\documentstyle[prl,aps,%
        epsf
]{revtex}
\begin{document}
\advance\textheight by 0.5in
\advance\topmargin by -0.25in
\draft

\twocolumn[\hsize\textwidth\columnwidth\hsize\csname@twocolumnfalse%
\endcsname

\preprint{SU-ITP \#95-3, cond-mat/9502060}

\title{ {\hfill\normalsize SU-ITP \#95-03, cond-mat/9502060\medskip\\}
Theory of the Resonant Neutron Scattering of High $T_c$
Superconductors}

\author{Eugene Demler and Shou-Cheng Zhang}
\address{ Department of Physics, Stanford University, Stanford, CA
  94305}

\date{December 20, 1994}

\maketitle

\begin{abstract}
  Recent polarized neutron scattering experiments on $YBa_2 Cu_3 O_7$
have revealed a sharp spectral peak at the $(\pi,\pi)$ in reciprocal
lattice centered around the energy transfer of $41\ meV$. We offer a
theoretical explanation of this remarkable experiment in terms of a
new collective mode in the particle
particle channel of the Hubbard model. This collective mode yields
valuable information about the symmetry of the superconducting gap.
\end{abstract}
\pacs{PACS numbers: 74.72.Bk, 61.12.Bt, 61.12.Ex}
]

Recently, both unpolarized and polarized neutron scat\-tering
experiments have been performed on the $YBa_2 Cu_3 O_7$ high $T_c$
superconductors \cite{grenoble,mook,keimer}. In particular, polarized
neutron experiment \cite{mook} shows an extremely sharp spectral
feature in the spin flip channel.  This feature is centered around
$(\pi,\pi)$ in reciprocal space, and peaked at $41\ meV$ with a width
narrower than the instrumental resolution.  This feature also has an
interesting temperature dependence.  In the experiment by Mook {\it et
  al}, while it exists above the superconducting transition
temperature of $T_c=92.4K$, its intensity scales like the superfluid
density below the transition.  More recently, Fong {\it et al}
performed detailed spin unpolarized neutron experiments with a careful
subtraction of the phonon background. They found that the $41\ meV$
mode disappears above the superconducting transition temperature.

In this letter, we offer a theoretical explanation of this
remarkable
experiment. We first show that for a general class of tight
binding
Hamiltonian
including the Hubbard and the $t-J$ model, there exist
well-defined collective
modes in the particle particle channel centered around
momentum
$(\pi,\pi)$. The spin quantum number of this excitation
can either be
a singlet or a triplet.
The singlet excitation has been discussed by Yang\cite{yang}
and one of
us\cite{zhang}, and is in fact an exact eigenstate of the
Hubbard model.
Normally, collective excitations in the particle particle
channel are
inaccessible experimentally. However, one of us\cite{zhang}
argued that if the ground state of the model
in consideration is superconducting, one can couple to it
through a
particle
hole excitation, because the BCS condensate is a coherent mixture
of particles and holes.
Based on this consideration, one of us\cite{zhang}
predicted a possible new collective mode of
the high $T_c$ superconductors. It is a spin singlet excitation
peaked at $(\pi,\pi)$, has a well
defined energy of $U-2\mu$ and its intensity scales like the
superfluid
density.
Possibly because it is hard to distinguish it from other
excitations in the
system, this mode has not yet been detected
experimentally.

However, the basic arguments can be easily generalized from the
singlet to
the triplet case. Besides the above mentioned collective mode
in the singlet
particle particle channel,
there also exists a well defined collective mode in the triplet
channel near
total momentum $(\pi,\pi)$. This is true for a large class of
tight binding
models, like the Hubbard or the $t-J$ model. The energy
spectrum of
a non-interacting pair of particles or holes generally
consists of a
continuum labeled by their relative momentum. However,
for tight binding
models, this continuum collapses to a point when the total
momentum of
the pair is  $(\pi,\pi)$. (See Figure 1).

\begin{figure}[hbt]
\epsfxsize=\columnwidth\epsfbox{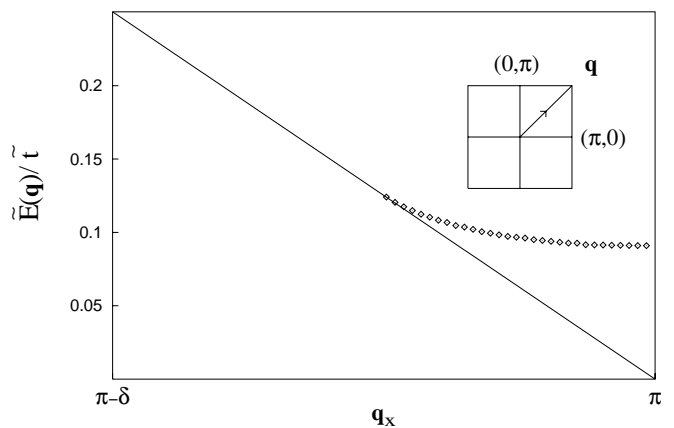}
\vspace{15pt}
\caption{ The energy spectrum along the $(\pi,\pi)$ direction. The
  dots correspond to the antibonding state and the solid line is
  the edge of the continuous spectrum. Here
  the numerical calculations were done for
  $J=\tilde{t}$ and $ n=0.85 $. As $\delta$ we denoted
  $\frac{\pi}{50}$. }
\label{dispersion}
\end{figure}

A triplet pair generally has a repulsive
interaction when placed on the neighboring site.
Because of the collapse of the particle particle continuum,
this repulsive interaction
leads to an {\it anti-bound} state near total momentum
$(\pi,\pi, ...)$
{\it in any space dimensions}. (See Figure 1).
This anti-bound triplet state manifests itself
as a collective excitation of the many-body system, or as
a pole in the
Greens function of the particle particle channel. Most
physical probes
do not couple to this channel. However, if the model in
consideration
is superconducting (which is an assumption of our theory),
then a spin flip scattering of the neutron
can couple directly to the triplet particle particle excitation.
Physically
this process is nothing but the spin flip scattering of a
Cooper pair
in the RCS condensate. There is an overwhelming evidence
that the Cooper
pair in the high $T_c$ materials is a spin singlet with
total momentum zero\cite{scalapino}.
Therefore a spin flip scattering of the Cooper pair
creates a triplet state. The
orbital angular momentum or the parity selection rule
forbids such a
transition if the momentum loss of the neutron is zero, on
the other hand, it can be shown
simply that the matrix element of this coupling is maximal
when the
momentum loss of the neutron is at $(\pi,\pi)$.

Based on the above reasoning, we shall interpret the sharp
spectral feature
observed in the polarized neutron experiment in terms of
the triplet collective mode in
the particle particle channel. We begin with some analytic
calculations
of the properties of this collective mode and its contribution
to the dynamical
spin spin correlation function. Subsequently, we compare our
predictions
with a number of characteristic features of the resonance
observed in experiments.
Finally we point out some difficulties with our interpretation
and their
possible resolution.

Our anti-bound state is different from the excitons of a
superconductor considered
by Bardasis and Schrieffer\cite{bardasis}. These excitons
form because
of an {\it attractive} potential in a given angular
momentum channel,
and they exists inside the superconducting gap and
near total
momentum $q=0$. Our model is also different from the
magnetic
susceptibility in the superconducting state computed
using
RPA\cite{monthoux}, since it involves multiple scattering
in the
particle hole channel. More recently, Bulut and
Scalapino\cite{bulut} considered
a model of the bi-layer superconductors and argued that
a dynamic
nesting from the bonding to anti-bonding Fermi surface
could give rise
to a collective resonance at $(\pi,\pi)$. Our main
difference
lies in the fact that their resonance is in the
particle hole channel
and as such it can exist well above $T_c$. The bi-layer
band structure
plays a crucial role in their model while it is irrelevant
in our case.

We consider the following model defined on a two dimensional square
lattice:
\begin{eqnarray} {\cal H} =
  &-&t \sum_{<ij>}{{c^{\dagger}_{i\sigma} c_{j\sigma}}}
  +J \sum_{<ij>}{{ \vec{S}_{i}\cdot \vec{S}_{j}}}\nonumber\\
  &+& U \sum_{i}{{ n_{i\uparrow} n_{i\downarrow}}}
  -\mu \sum_{i}{{c^{\dagger}_{i\sigma} c_{i\sigma}}}
\label{model}
\end{eqnarray}
In the limit $U\rightarrow\infty$, we recover the $t-J$ model.
On the
other hand, if we keep $U$ small, we can regard the above model
as an effective Hamiltonian of the weak coupling Hubbard model,
where the
$J$ term arises from the paramagnon interaction on the nearest
neighbor sites. To study the spectrum in the triplet particle
particle channel, we consider the operator
\begin{equation}
O^\dagger_{q}= \sum_{p} f_{q}\left(p\right) c_{p+q\uparrow}^\dagger
c_{-p\uparrow}^\dagger \end{equation}
which describes a pair of particles with center of mass momentum
$q$ and relative momentum $p$. The equation of motion for this
operator is given by its commutator with the Hamiltonian,
$\left[{\cal H}, O^\dagger_{q}\right]$. If we factorize the
resulting
commutator in terms of  $O^\dagger_{q}$ and the expectation
values
of the density $n_{k\sigma}$, we obtain:
\begin{eqnarray} \lefteqn
 {\left[{\cal H}, O^\dagger_{q}\right]    =}
 \nonumber \hspace{0.7in} \\
\lefteqn { \sum_{p} (\epsilon_{p+q} + \epsilon_{-p} ) f_{q}(p)
  c^\dagger_{p+q\uparrow} c^\dagger_{-p\uparrow}
+  2 (U n_{\downarrow}- \mu) O^\dagger_{q}}
 \nonumber \hspace{.5in}\\
\lefteqn { \hspace{-13mm}+ \frac{J}{8} \sum_{p}
{{c^\dagger_{p+q\uparrow}
c^\dagger_{-p\uparrow}}}\times  \frac {1}{N} \sum_{p'} f_{q}(p')}
\nonumber \\
\lefteqn{  \hspace{-4mm}\times \left(
\eta(p-p')-\eta(p+p'+q) \right)(1-n_{p'-q\uparrow}-n_{p'\uparrow})}
\nonumber\\
\lefteqn {-\frac{3J}{8} \sum_{p} f_{q}(p) {{c^\dagger_{p+q\uparrow}
c^\dagger_{-p\uparrow}}}}
 \nonumber \hspace{.5in} \\
& \times \frac {1}{N} \sum_{p'}& n_{p'}
\left( \eta(p+p')+\eta(p'-p-q) \right)
\end{eqnarray}
where $n_{\sigma}=\frac{1}{N}\sum_k n_{k\sigma}$ is the
average density
of electrons with spin $\sigma$, $N$ is the
total number of lattice sites and and
\mbox{$\eta(p)=\sum_{\vec{a}}\exp[i\vec{p}\cdot\vec{a}] $} is a
geometrical factor
coming from the summation over the nearest neighbors.
In this equation, the first term describes the kinetic energy of
the pair of particles in consideration. For tight binding models
with nearest neighbor hopping, $\epsilon_k=-2t(\cos k_x+\cos k_y)$.
In this case, the kinetic energy of the pair vanishes when the
total momentum $q=(\pi,\pi)$. The second term describes the
Hartree interaction of the spin up pair with the average
density of the
down spins in the background and the chemical potential energy
of adding a pair of particles. In the large $U$ limit, the
leading contribution to the chemical potential is given by
$U n/2$. In fact, for Hubbard model at half filling, {\it i.e.}
at $n=1$, $\mu=U/2$ is an exact relation. Therefore, the
second terms cancels in the leading order in $U$, and reaches
a finite limit as $U\rightarrow\infty$.
The third term gives the multiple scattering of the
particles with each
other on a restricted phase space due to the filled
Fermi sea.
The last term describes the Fock self-energy of the
quasi-particles.
The method of factorizing the operator equation of motion is
equivalent to the $T$ matrix approximation in the diagrammatic
calculations, it is exact in the low density limit.
The collective mode for the triplet particle particle excitation
is obtained by equating the right hand side of (3)
to $E_q O^\dagger_{q}$, the resulting eigenvalue equation is given
by
\begin{eqnarray}
  1\!=\!\frac{J}{2N}\!\!\sum_{p}\! \frac
  {\sin^{2}(p_{\alpha})(1-n_{{\scriptstyle q\over \scriptstyle 2}-
p\uparrow}-
    n_{{\scriptstyle q\over \scriptstyle
        2}+p\uparrow})} { \tilde E_{q}^{\alpha}\!+\!4\tilde{t}
\left[
    \cos(p_{x})\cos(\frac{q_{x}}{2})\!+\!\cos(p_{y})
\cos(\frac{q_{y}}{2})
  \right] }
\label{eigenvalue}
\end{eqnarray}
with $\alpha $ being $x$ or $y$, \mbox{ $ \tilde{t}=
t+\frac{3J}{4N}
  \sum_{p} n_{p\downarrow}
 \cos(p_{x}) $} and \mbox{$E_q=\tilde{E_q}+ 2(Un_\downarrow-\mu)$}.
It is straightforward to see that a discrete eigenvalue exists
at $q=(\pi,\pi)$, with energy of
\begin{equation}
\tilde E=\frac{J}{4}(1- \frac{2}{N}\sum_p n_{p}
\cos^2 p_x)
\label{En}
\end{equation}
Discrete
eigenvalues also exist for a finite range in $q$ space. There are
in general two eigenvalues for each $q$, one corresponds to
a $p$ wave pair oriented in the $x$ direction, while the
other is oriented in the $y$ direction. The two eigenvalues
are degenerate only along the diagonal when $q_x=q_y$. The above
eigenvalue equation can easily be solved numerically and
the dispersion
of the collective mode is shown in figure 1.  $\tilde{t}$
is basically
the renormalized hopping matrix element of the quasi-particle.
The range
in momentum space over which the collective mode exists depends
on the
ratio of $J/\tilde{t}$. Numerous numerical calculations
indicate that
the quasi-particle bandwidth is of the order of
$J$\cite{dagotto}.
Here we take a semi-phenomenological approach and
choose $\tilde t$
rather than $t$ as a free parameter. {}For a ratio of
$J/\tilde{t}=1$, and $n=0.85$ we see that the collective
mode exists
over $\frac{\pi}{50}$ of the momentum space.

In the experiments with the polarized neutrons scattering one
measures the dynamic spin-spin correlation function
\begin{equation} S(q,\omega)= \sum_{n} \mid
\langle n \mid S_{q} ^\dagger \mid  0\rangle \mid ^{2}
\delta(\omega-\omega_{n0})
\end{equation}
where $\mid 0>$ and $\mid n>$ are the ground and excited
states of the
system and
$S^\dagger_{q}=\sum_{p}c^\dagger_{p+q\uparrow}c_{p\downarrow}$.
Using the operator equation
$\left[{\cal H}, O^\dagger_{q,\alpha}\right] = E_{q}^{\alpha}
O^\dagger_
{q,\alpha}$
we can construct a class of approximate
excited states of the Hubbard Hamiltonian as
$\mid n\rangle=O^\dagger_{q, \alpha}\mid 0\rangle$. The same
operator equation for
$O_{q,\alpha}$ shows that $O_{q,\alpha}\mid 0\rangle =0$.
With these relations, we can calculate the contribution of this
approximate eigenstate to $S(q,\omega)$ at zero temperature:
\begin{eqnarray} &S_{0}&(q,\omega)=\nonumber\\
&~~&=\sum_{\alpha}\langle 0 \mid  O_{q,\alpha}
 S^\dagger_{q} \mid 0\rangle \mid ^{2}
 \delta(\omega-E_{q}^{\alpha}) + \sum_{n'} \nonumber\\&~~&=
 \sum_{\alpha}\langle 0 \mid \left[ O_{q,\alpha},
 S^\dagger_{q}\right] \mid 0\rangle \mid ^{2} \delta(\omega-E_{q}^
{\alpha})+\sum_{n'}
\end{eqnarray}
where $\sum_{n'}$ denotes the contribution from states other than
$O^\dagger_{q, \alpha}\mid 0\rangle $, and
we used the fact that $O_{q, \alpha}\mid 0\rangle =0$ to replace the
product of two
operators by their commutator. Evaluating the commutator we obtain
\begin{eqnarray} \lefteqn{ S_{0}(q,\omega)=} \nonumber
\hspace{.0in}\\
&&\sum_{\alpha} \mid\langle 0 \mid 2 \sum_{p}
  f_{q}^{(\alpha)}(p-\frac{q}{2}) c_{-p+\frac{q}{2}\uparrow}
  c_{p-\frac{q}{2}\downarrow}\mid 0\rangle \mid ^{2}
  \delta(\omega\!-\! E_{q}^{\alpha})\nonumber\\
&&=4\sum_{\alpha} \mid \sum_{p} f^{(\alpha)}_{q}(p-\frac{q}{2})
\Delta_{p-q/2} \mid^2
\delta(\omega-E_{q}^{\alpha})
\end{eqnarray}
Summation over $\alpha$ corresponds to the two kinds of
antibonding
solutions. The overlap matrix element is finite for both
extended
$s$ wave and $d$ wave pairing in the ground state.
If we take the $d$ wave symmetry of the order parameter
$\Delta_{k}=\Delta_{0}(\cos(k_{x})-\cos(k_{y}))$ and, for example,
$f_{q}^{(x)}$ - a $p$ wave oriented in the $x$ direction. Then,
\begin{eqnarray} &&\sum_{p}f_{q}^{(x)}(p-\frac{q}{2})
\Delta_{p-\frac{q}{2}}=
  \nonumber\\&=&
 \frac{1}{A_{q}}\sum_{p}
 \frac{\sin(p_{x})\Delta_{p-\frac{q}{2}}
(1-n_{{\scriptstyle q\over \scriptstyle 2}-p}-n_{{\scriptstyle
q\over
\scriptstyle 2}+p}) }
{\tilde E_{q}^{x}+4\tilde{t}(\cos (p_{x}) \cos(
    \frac{q_x}{2})+\cos (p_{y}) \cos(\frac{q_y}{2}))}=
\nonumber\\&=&
-\frac{2 \Delta_{0}}{JA_{q}}\sin(\frac{q_{x}}{2})
\end{eqnarray}
where $A_{q}$ is the normalization factor of the wave function
\begin{eqnarray*}
A_{q}^{2}= \sum_{p} \frac{\sin^{2}(p_{x})(1-n_{{\scriptstyle
q\over
\scriptstyle 2}-p}-n_{{\scriptstyle q\over
\scriptstyle 2}+p})^2 }{
\left[ \tilde E_{q}+4\tilde{t}(\cos (p_{x}) \cos(\frac{q_x}{2})+
\cos (p_{y})
\cos(\frac{q_y}{2}))\right]^{2}}
\end{eqnarray*}
Finally we have
\begin{eqnarray}  S(q,\omega,T)=16 \frac{\Delta_{0}^{2}(T)}
{J^{2}A_{q}^{2}}
[\sin^{2} (\frac{q_{x}}{2}) \delta(\omega-E_{q}^{x})
\nonumber \\ +
\sin^{2} (\frac{q_{y}}{2}) \delta(\omega-E_{q}^{y})]
\left(1+n_B(\frac{\omega}{T})\right)
\end{eqnarray}
where the last factor takes care of the finite temperature in
spin-spin correlation functions. One can easily see though,
that the
main temperature dependence of $ S(q,\omega,T)$ comes from
$\Delta_{0}(T)$. We also see that the intensity vanishes at
zero external momentum and is maximal at $(\pi,\pi)$. This fact
can simply be attributed to the conservation of orbital angular
momentum. Since the Cooper pair in the ground state has
angular momentum either $l=0$ or $l=2$, a finite momentum transfer
is required for it to be excited to a state with angular momentum
$l=1$. The dispersion of two states is not very big giving
rise to only a small splitting of two levels which never exceeds
$15\%$.

We also see that the spectral intensity is simply proportional
to the BCS order parameter or the superfluid density. This is
so because the BCS order parameter provides the coupling from
the particle hole channel to the particle particle channel
\cite{zhang}.
This extra spectral weight at energy $E_{q,\alpha}$ is transfered
from the low energy sector, since the singlet BCS pairing
removes the low energy spin fluctuation. Our theory
is consistent with the fact that the $41 meV$ peak intensity seems
to scale with the superfluid density below $T_c$.
This conclusion agrees with the experimental results obtained by
Fong {\it et al}\cite{keimer},
but it is inconsistent with the fact that this mode does not
vanish above $T_c$ in the experiment by Mook {\it et al}\cite{mook}.
At this moment, it is an experimental question that has to be
resolved.
We also note that our calculation is not fully self-consistent,
and
assumes that the mode energy does not change significantly
when the
system becomes superconducting. This is true for a variety
of collective
modes like the plasmon and the singlet particle particle mode
\cite{yang,zhang} (the so called $\eta$ pair). But we do
not have a
general
proof in the present case. A fully self-consistent calculation will
be carried out in the future\cite{collaboration},
but we anticipate
that the basic physics remain unchanged.

All the above discussions were restricted to the two
dimensional
$Cu O$ plane, which we model by the two dimensional
Hubbard model.
However, it can be simply generalized to three dimensions.
If one takes
a three dimensional Hubbard model, one finds that collective
mode exists near $(\pi,\pi,\pi)$ rather than $(\pi,\pi,0)$.
This
is consistent with the experiment where the third component of the
momentum transfer is also $\pi$.

The fact that the particle particle collective mode always
exists at
$(\pi,\pi,..)$ is a special property of the tight binding
model on a
bi-partite
lattice. How would a next-nearest neighbor hopping term
change the results?
In this case, the dispersion relation is given by
\begin{eqnarray*}
  \epsilon_k\!=\!-\!2\tilde{t}(\cos k_x\!+\!\cos
  k_y)\!-\!2t'(\cos(k_x\!+\!k_y)\!+\!\cos(k_x\!-\!k_y))
\end{eqnarray*}
where $t'$ denotes the amplitude of the next-nearest neighbor
hopping.
The $x$ and the $y$ collective modes are mixed
into the bonding $\sin(p_x) + \sin(p_y)$ and the antibonding
$\sin(p_x) - \sin(p_y)$ combination, and
we obtain a matrix equation for the eigenvalues of these
collective modes:
\begin{eqnarray}
det\parallel\frac{J}{2}\frac{1}{N}\sum_p \frac{\sin(p_\alpha)
\sin(p_\beta)
z_{pq}}{\tilde E-\Omega_{pq}}
-\delta_{\alpha \beta}\parallel=0
\label{t'}
\end{eqnarray}
where $
z_{pq}=1-n_{\frac{q}{2}-p}-n_{\frac{q}{2}+p}
$
and $
\Omega_{pq}=\epsilon_{\frac{q}{2}-p}+\epsilon_{\frac{q}{2}+p}$.
It is
easy to see that the antibonding state $E^{-}_{\pi,\pi}$ always
exists, while there is a critical
coupling
\begin{eqnarray}
t'_{cr}=1.2 \times 10^{-4} J.
\end{eqnarray}
{}for the bonding state,
so that for $t'>t'_{cr}$, it ceases to exist at $(\pi,\pi)$.
In real experiments $t'$ and $J$ are of the same order of
magnitude,
therefore, one can safely conclude that the anti-bonding
state disappears into the continuum. This conclusion leads
directly to a measurement of the {\it symmetry of the
superconducting
gap}. Below the superconducting transition temperature,
the intensity
of the anti-bonding state as measured in the neutron
scattering
experiment is proportional to
\begin{eqnarray}
\mid \sum_{p} f^{(\alpha)}_{q}(p-\frac{q}{2}) \Delta_{p-q/2}
\mid^2
\end{eqnarray}
where $f^{(\alpha)}_{q}(p-\frac{q}{2})\propto \sin(p_x)-
\sin(p_y)$. From this
equation we see immediately that the intensity is non vanishing
if and only if the gap symmetry is of the $d$ wave type.
We therefore
argue that the existence of the neutron resonance determines the
{\it symmetry} of the pairing gap of the high $T_c$ superconductors
to be of the $d$ wave type, consistent with the theories where
pairing interaction arises from the spin
fluctuations\cite{scalapino2,spinbag,pines}. It could also be
consistent with more exotic possibilities of $d_{x^2-y^2} +
i d_{xy}$ pairing symmetry\cite{laughlin}.

We conclude that the basic features of the observed polarized
neutron
scattering experiment can be explained in terms of a new
particle particle collective
mode in the Hubbard model. The energy of the mode at $(\pi,\pi)$
is given by (\ref{En}) in the case of $t'=0$ and implicitly by
(\ref{t'}) for finite $t'$. Unfortunately, the energy of the mode
depends on all three parameters $\mu$, $J$ and $t'$.
While these
parameters can be fitted to the experimental value, the fitting
itself is not a confirmation of our theory. The
mode is centered around $(\pi,\pi,\pi)$ for two different reasons,
both
because the particle particle continuum at this momentum is
minimal so that the anti-bound
state could exist, and because of the conservation of angular
momentum
for exciting a singlet Cooper pair to a triplet state. This is
exactly the
momentum transfer of the excitation observed in experiment.
The intensity of the mode scales with the superfluid density
because
the BCS pairing amplitude is involved in converting a particle
hole pair
into a particle particle pair. Above $T_c$, our simple calculation
shows
that the mode disappears, a much more sophisticated calculation
is required
to understand if the fluctuation pairing effects could give
rise to a resonance even above $T_c$. When the effects of
next-nearest-neighbor
hopping is included, only the anti-bonding $\sin(p_x)-\sin(p_y)$
mode
can exist and the region in momentum space for the
existence of this collective mode becomes much narrower.
However, the
sharp collective mode might still exist as a broad
resonance outside
of this region. The antibonding collective mode only has
an overlap with
the $d$ wave order parameter, and
we conclude that the experimental observation
of the collective resonance in the neutron scattering
experiment can
only be consistent with the $d$ wave symmetry of the
pairing gap.

We are extremely grateful to Prof. D. Scalapino for a
stimulating
journal club talk on the experiment and generous
sharing of his
ideas on the problem. We would also like to thank
Prof. R. B.
Laughlin, V. Emery, N. Bulut and Z. X. Shen
for many enlightening discussion on the
experiment. Part of this work is supported by the Center for
Materials Research at Stanford University.


\begin{references}

\bibitem{grenoble} J.~ Rossat-Mignod, {\it et al},
{\em Physica} {\bf 180B}, 383 (1992).

\bibitem{mook} H.~A.~ Mook, {\it et al},
{\em Phys. Rev. Lett.} {\bf 70}, 3490 (1993).

\bibitem{keimer} H. F. Fong {it et al},
Princeton University preprint.

\bibitem{yang} C.~N.~ Yang, {\em Phys. Rev. Lett.} {\bf 63},
2144 (1989);  C.N. Yang and S.C. Zhang,
{\em Mod. Phys. Lett.} {\bf B4}, 759 (1990).

\bibitem{zhang} S~.C.~ Zhang, {\em Phys. Rev. Lett.} {\bf 65},
120 (1990); S. C. Zhang, {\em Int. J. Mod. Phys.} {\bf B5},
153 (1991).

\bibitem{scalapino} D. J. Scalapino, {\em Phys. Rep.}
to be published.
{\em Phys. Rev.} {\bf 121}, 1050, (1961).

\bibitem{bardasis} A. Bardasis and J. R. Schrieefer,
{\em Phys. Rev.} {\bf 121}, 1050, (1961).

\bibitem{monthoux} P. Monthoux and D. J. Scalapino,
{\em Phys. Rev. Lett.} {\bf 72},
1874 (1994).

\bibitem{bulut} N. Bulut and D. J. Scalapino, UCSB preprint.

\bibitem{dagotto} E. Dagotto, {\em Rev. Mod. Phys.}
{\bf 66}, 763 (1994).

\bibitem{collaboration} N. Bulut, E. Demler, D. J. Scalapino
and S. C. Zhang,
work in progress.

\bibitem{scalapino2} D.~J.~Scalapino, J.~E.~Hirsch and E.~Y.~Loh,
{\em Phys. Rev.} {\bf B 34}, 8190 (1986).

\bibitem{spinbag} J.~R.~Schrieffer, X.~G.~Wen and S.~C.~Zhang,
{\em Phys. Rev.} {\bf B 41}, 6399 (1990).

\bibitem{pines} P.~Monthoux, A.~V.~Balatsky and D.~Pines,
{\em Phys. Rev.} {\bf B46}, 14803 (1992).

\bibitem{laughlin} R.~B.~Laughlin, Stanford University preprint.

\end{references}
\end{document}